\documentclass[12pt]{iopart}

\usepackage{graphicx}
\begin{document}

\title[Intermediate-Mass Black Holes as LISA Sources]{Intermediate-Mass
Black Holes as LISA Sources}

\author{M. Coleman Miller}

\address{University of Maryland, Department of Astronomy and
Maryland Astronomy Center for Theory and Computation, College
Park, MD 20742-2421, USA}
\ead{miller@astro.umd.edu}
\begin{abstract}

Intermediate-mass black holes (IMBHs), with masses in the range
$\sim 10^{2-4}~M_\odot$, will be unique sources of gravitational
waves for LISA.  Here we discuss their context as well as
specific characteristics of IMBH-IMBH and IMBH-supermassive
black hole mergers and how these would allow sensitive tests
of the predictions of general relativity in strong gravity.

\end{abstract}

\maketitle

\section{Introduction}

Conclusive evidence exists for black holes in the stellar-mass
($\sim 3-30~M_\odot$) and supermassive (SMBH; $\sim
10^6-10^{10}~M_\odot$) ranges.  This evidence comes from direct
measurements of masses from orbital motion in, respectively,
binaries and galactic nuclei.  In contrast, the range between
these masses does not yet provide clear dynamical evidence for
black holes of intermediate mass (IMBHs).  This is essentially
because candidates are rare (hence the nearest possible IMBH in
a binary is not easily observed) and have small radii of
influence (hence SMBH-like observations of nearby stars are also
challenging).  There are, however, numerous indirect suggestions
of the $\sim 10^{2-4}~M_\odot$ mass range from the high fluxes
of ultraluminous X-ray sources (ULXs; see \cite{Fab89,CM99}),
from their relatively low thermal temperatures
\cite{M03,M04a,M04b}, and from aspects of the dynamics of
globular clusters \cite{vdM02,GRH02,NGB08}.  See
\cite{MillerColbert04} for a recent summary of the data.  The
lack of definitive dynamical evidence means that alternate
scenarios have also been proposed for ULXs, including beamed
emission \cite{Rey97,Ki01} and super-Eddington emission
\cite{RB03,Beg06}.

Here we assume that IMBHs exist and explore various ways in
which their interactions could lead to gravitational radiation
detectable in the $\sim 10^{-4}-10^{-2}$~Hz frequency range
of the {\it Laser Interferometer Space Antenna} (LISA).
In \S~2 we discuss the broader context of IMBHs and proposed
formation mechanisms.  We also give basic formulae for the
amplitude and frequency of gravitational waves from binaries,
showing why IMBHs could be bridge sources between space-based
and ground-based detectors.  In \S~3.1 we evaluate the prospects
for IMBH-IMBH mergers and what they might tell us.  In \S~3.2 we
discuss IMBH-SMBH mergers and their prospects as uniquely
precise testbeds for strong gravity predictions of general
relativity.  We present our conclusions in \S~4.

\section{Context and formation of IMBHs}

Most formation scenarios of supermassive black holes propose
that they pass through a life stage in which they are IMBHs.
Therefore, study of IMBHs can yield insight into the early
formation of structure in the universe.  This is particularly
true if the formation of SMBH seeds is in part due to interactions
in dense stellar clusters, as this is a leading candidate for
how IMBHs form in the current universe.  We now discuss proposed
formation mechanisms and their implications.

\subsection{Suggested formation mechanisms for IMBHs}

The reason for setting the lower limit on IMBHs at $\sim 100~M_\odot$
is that this appears to be in excess of the maximum black hole mass
that can form from a solitary star in the current universe 
(see \cite{MR01} for a discussion in the context of IMBHs).
IMBH existence therefore requires new formation scenarios.
The three basic ideas are:

\begin{itemize}

\item The first generation of stars had negligible metallicity.
This reduces radiative opacity and thus the strength of radiation
driven winds, hence stars can start more massive and retain more
of their mass than current stars \cite{MR01}.

\item In young massive stellar clusters that have relaxation times
for the most massive stars that are less than the lifetimes of those
stars (about 2.5~million years), mass segregation leads to collisions
and mergers of massive stars.  If the combined objects do not have
catastrophic wind mass loss (see \cite{Belkus07}), they can in principle
accumulate up to thousands of solar masses and might then become
black holes with similar masses \cite{E01,PZ02,PZ04,G04}.

\item If a seed black hole with a mass of more than $\sim 200~M_\odot$
is formed by the previous process in a dense stellar cluster, subsequent
binary-single and binary-binary interactions can allow it to merge
and accumulate mass without being vulnerable to ejection by either
three-body kicks or recoil from asymmetric gravitational wave
emission \cite{MH02a,MH02b,Gultekin2004,Gultekin2006,OLeary2006,OLeary2007}.

\end{itemize}

Of special interest to LISA observations is that some simulations
suggest that if the initial binary fraction in a stellar cluster
exceeds $\sim 10$\% then more than one IMBH can form in that cluster
(\cite{G04}; but see \cite{Belkus07}).
We will explore the consequences of this in \S~3.1.
In addition, since massive stellar clusters are often found near
the nuclei of galaxies that are interacting actively, the clusters
themselves can sink to the centers of galaxies where their IMBHs
spiral into the galaxy's supermassive black hole.  We discuss this
in \S~3.2.

The last piece of basic physics has to do with gravitational
waves themselves.  For a circular binary with a total mass $M=m_1+m_2$ 
and a reduced mass of $\mu=m_1m_2/M$ at a distance $d$ from us small
enough that redshifts are unimportant,
orbiting at a frequency $f_{\rm orb}$ so that the dominant gravitational
wave frequency is $f_{\rm GW}=2f_{\rm orb}$, the angle-averaged
dimensionless strain amplitude that we observe is
\begin{equation}
h=6\times 10^{-21}\left(f_{\rm GW}/1~{\rm Hz}\right)^{2/3}
\left(M_{\rm ch}/10^3~M_\odot\right)^{5/3}\left(1~{\rm Gpc}/d\right)\; .
\end{equation}
Here $M_{\rm ch}$ is the ``chirp mass", defined as
$M^{5/3}_{\rm ch}\equiv \mu M^{2/3}$.  The maximum frequency
of orbits that evolve relatively slowly is often approximated
by the frequency at the innermost stable circular orbit (ISCO),
although technically the ISCO concept is strictly valid only
when there are no mechanisms for angular momentum loss (i.e.,
for test particles in geodesic orbits).  For a nonrotating
spacetimes, this maximum frequency is
\begin{equation}
f_{\rm GW,max}({\rm ISCO})=4.4~{\rm Hz}(10^3 M_\odot/M)\; .
\end{equation}
From this expression we see that IMBHs in the entire mass
range of $M\sim 10^{2-4}~M_\odot$ are potential LISA sources,
but also that towards the low end of the masses they might
be sources for ground-based detectors, which focus on
$f_{\rm GW}>10~{\rm Hz}$.

\section{Gravitational waves from mergers of IMBHs with other black holes}

We will focus on mergers of IMBHs with other IMBHs or with SMBHs,
because the rate at which LISA will detect the coalescence of
stellar-mass black holes with IMBHs is negligibly low \cite{Will04},
although such mergers may be detected by next-generation ground-based
instruments such as Advanced LIGO or Advanced Virgo \cite{Man08}.

\subsection{IMBH-IMBH mergers}

As demonstrated by \cite{Fre06}, if more than one IMBH forms
in a young stellar cluster, as may happen if the primordial binary
fraction in the cluster exceeds $\sim$10\% \cite{G04}, the subsequent
coalescence of the IMBHs can be visible out to large distances.
Specifically, \cite{Fre06} found that a comparable-mass binary with a
total rest-frame mass of $1000~M_\odot$ would have a coalescence
visible with LISA out to a redshift $z\approx 1$.  Given that
the star formation rate increases dramatically from $z=0$ to
$z=1$, if these mergers occur within a few tens of millions
of years after the formation of the clusters then LISA observations
of such events could be unique probes of star formation and
cluster dynamics.

To explore this further, \cite{Ama08} 
performed detailed N-body simulations of two IMBHs in a cluster.
They started with equal-mass IMBHs (either $300~M_\odot$ or
$1000~M_\odot$ each) at a separation of 0.1~pc, in a cluster
of $3.2\times 10^4$ stars either all at $1~M_\odot$ or selected
from a Kroupa \cite{Kroupa00b} initial mass function.  They then followed
the inspiral of the IMBHs until the black holes eventually formed a very
hard binary.  At that point, they passed off the properties of
the binary to a three-body scattering program, where the speeds
and masses of the interacting stars were drawn from the appropriate
external distribution.  Their three major conclusions are:

\begin{itemize}

\item In the simulations, and by extension in real clusters with
more stars, the IMBH merger has only a minor effect on the cluster
in general.  Superficially it might seem that the effect could be
substantial, because the binding energy of an IMBH binary at the
point of the last scattering with a star can exceed the total
binding energy of the cluster by a significant factor.   Given that
interactions with stars are what harden the binary to this point, it
might thus appear that the cluster could be disrupted. The reason
that this does not happen is that the energy extracted from
three-body binary hardening is only put back into the cluster if the
star emerges with a speed less than the escape speed of the
cluster.  In contrast, high-speed ejections share very little energy
with the cluster because the relaxation time (which is needed to
alter energies significantly) is orders of magnitude greater than
the time to leave the cluster.  Since the binding energy of the IMBH
binary is much less than the binding energy of the cluster at the
point when stars are ejected from the cluster, the net effect on the
cluster is small.

\item The duration of merger is indeed tens of millions of years
or less, and as expected is dominated by the phase in which the
binary is hard (because the cross section for interactions is less
than when the binary is wider).  In none of the dozens of runs done
in \cite{Ama08} did this phase take more than $10^8$~years.
As a result, these mergers will indeed serve as good snapshots of
star formation.

\item In all other categories of comparable-mass black hole mergers,
the expected eccentricity upon entry into the frequency band of the
relevant detector (LISA for supermassive black holes; Virgo or LIGO
for stellar-mass holes) is likely to be so small as to be basically 
negligible (see \cite{Ama07} for a discussion;
for potential exceptions for stellar-mass holes, see
\cite{Wen03} and \cite{OLeary2008}).  For IMBH-IMBH coalescences in young
clusters, however, \cite{Ama08} find that at the LISA
lower frequency limit of $f_{\rm GW}\approx 10^{-4}$~Hz the eccentricity
is commonly $e\sim 0.1-0.3$.  As they demonstrate, this is detectable
and will have to be included in algorithms that characterize the
inspiral.

\end{itemize}

In Figure~1 we show two sample evolutions of
eccentricity versus semimajor axis for runs similar to those of
\cite{Ama08}.  The systematic increase in eccentricity
down to separations of $a\sim$few AU is consistent with the results
of \cite{Quinlan96}, who found that a massive binary interacting with
much less massive stars commonly increases its eccentricity.  The
decrease at smaller semimajor axes is the result of circularization
by gravitational radiation dominating over increases in eccentricity
caused by three-body interactions.

\begin{figure}
\resizebox{\hsize}{!}{\includegraphics{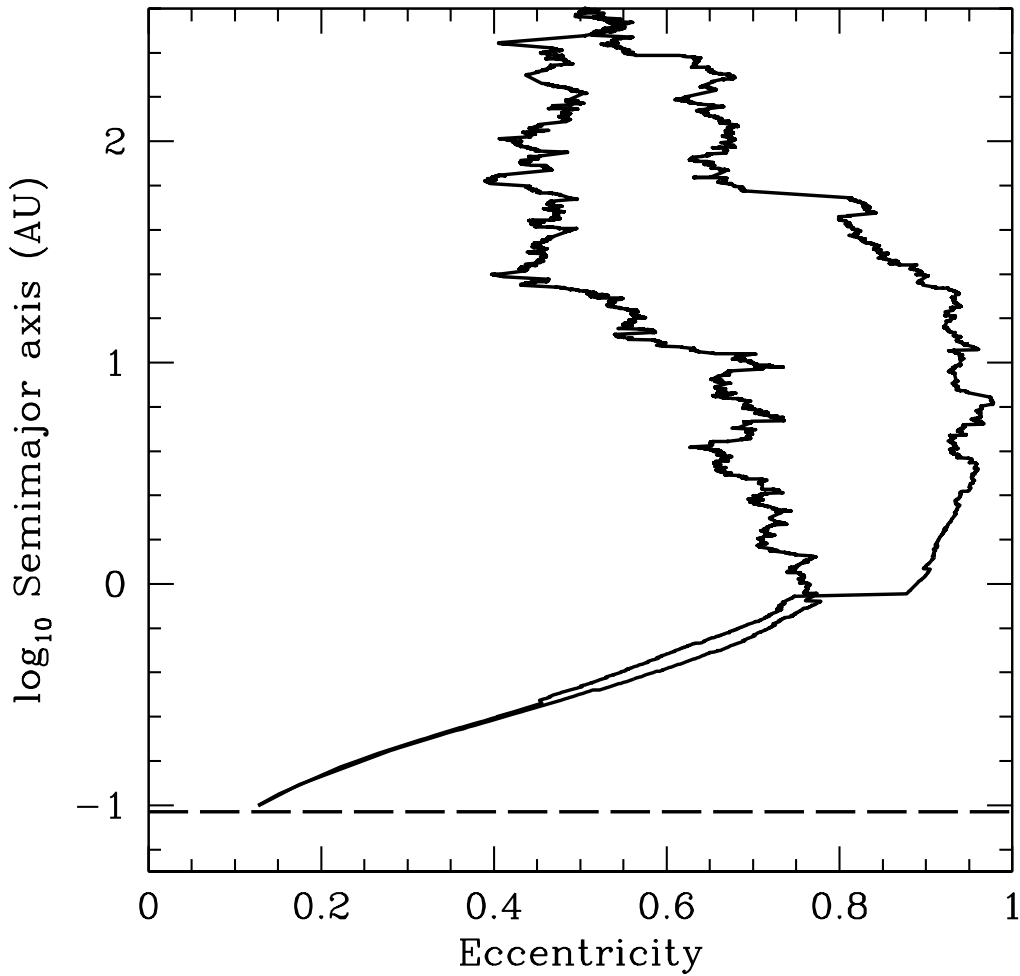}}
\caption{Two sample tracks (black lines) in semimajor axis versus eccentricity
for a $1000~M_\odot-1000~M_\odot$ binary starting at $a=400$~AU
and $e=0.5$ in a stellar cluster of total number density
$n=3\times 10^5$~pc$^{-3}$ with a Kroupa \cite{Kroupa00b} mass
function.  The dashed line shows the semimajor axis at which the
dominant gravitational wave frequency is $f_{\rm GW}=10^{-4}$~Hz,
the lower end of the LISA sensitivity band.
The convergence at small semimajor axis occurs because
for these cluster and binary parameters inspiral becomes dominated 
by gravitational radiation when the pericenter distance becomes
less than about 0.1~AU.  This figure is similar to figures in
\cite{Ama08}.\label{fig:binary}}
\end{figure}

\subsection{IMBH-SMBH Mergers}

When IMBHs merge with the central supermassive black holes in
galaxies, the resulting gravitational waveform contains uniquely
precise information about the spacetime around a rotating black
hole.  The basic reason is that the mass ratio is extreme enough
(typically $\sim 1000:1$) that approximate techniques can be used
to follow the inspiral, without the need to resort to
computationally expensive numerical relativity simulations.  These
are therefore similar to the more familiar extreme mass ratio
inspirals (EMRIs), in which a stellar-mass black hole of typically
$\sim 10~M_\odot$ coalesces with a supermassive black hole
(see \cite{Ama07} for a review). 
However, when the secondary is an IMBH with a mass of $\sim
1000~M_\odot$ the amplitude of the waves at a given luminosity
distance is two orders of magnitude greater than for a standard
EMRI, hence the waveform and comparisons of it with general
relativistic predictions can be obtained with far greater
precision. 

\cite{M05} showed that for favorable cases (e.g.,
$M_{\rm SMBH}=10^{5.5}~M_\odot$, $M_{\rm IMBH}=10^3~M_\odot$,
and an angular diameter distance of 3~Gpc) the signal to noise
near the end of the inspiral can be great enough that the source
would be detectable with LISA in a standard power density 
spectrum, without the need for matched filtering.  More specifically,
if one took a power density spectrum over the optimal period of
time, in which the frequency drift during this period is equal to
the frequency resolution of the spectrum (which equals the reciprocal
of the period of integration), then the signal to noise of the
single peak occupied by the inspiral during this period would
be $S/N=$tens.  As a result, it would be possible to string together
such detections and connect the phases of the inspiral, thus
building up an empirical waveform without the need to assume
that general relativity is correct.  Hence even a single detection
of such an event would provide a uniquely powerful test of the
properties of massive spinning black holes.

The rate of IMBH-SMBH mergers depends on various details of the
dynamics of IMBHs in dense stellar clusters.  The basic idea is
that although IMBHs not in galactic centers cannot by themselves
spiral to the core within a Hubble time (because dynamical friction
on such light objects is too weak), if they form in massive
clusters within tens to possibly hundreds of parsecs of the center
then the cluster will sink as a unit within a few billion years. 
The cluster itself is eventually  disrupted by the tidal field of
the galaxy and supermassive black hole, leaving the now solitary
IMBH much closer than before and able, in principle, to spiral in
to merger (see \cite{M05}).  This has been proposed as one
mechanism to shepherd the young, massive S stars observed near the
center of our Galaxy \cite{Han03}.  \cite{PZ06,Mat07} examined this
process using N-body simulations, and concluded that the rate of
LISA detections of these events could be tens per year, depending
on how efficiently IMBHs form in clusters.

More recently, \cite{Koch08} explored
additional effects, such as the interactions of IMBHs with themselves
around an SMBH, assuming that the full coalescence process takes
longer than the time needed for new clusters and IMBHs to sink to
the center.  They found that such encounters tend to eject one IMBH
(although slowly, so that it will sink back in), and leave the
other in an eccentric orbit that decays readily by emission of
gravitational radiation.  Regardless of the properties or rates
of such encounters, any detected by LISA will be valuable probes
of strong gravity.

\section{Conclusions}

The evidence for IMBHs is currently strong but circumstantial,
pointing to the need for dynamical mass measurements of binary
motion that will establish their existence definitively.  
Nonetheless, their likely formation mechanisms and dynamical
interactions link them to many exciting topics in the current
and early universe.  As gravitational wave sources they will be
unique in several respects: as bridge objects between space-based
and ground-based detectors, as comparable-mass binaries with
palpable eccentricities, and as the events that potentially will yield
the most precise tests of general relativity.  Many explorations
need to be done, but current results are highly encouraging for
their study.

\ack

We thank Tal Alexander, Pau Amaro-Seoane, Mike Gill, Doug Hamilton, 
Clovis Hopman, Vanessa Lauburg, Fred Rasio, Derek Richardson, and
Michele Trenti for stimulating conversations.  This work was supported
in part by NASA ATFP grant NNX08AH29G.

\end{document}